\documentclass{article}
\usepackage[english]{babel}
\usepackage[utf8]{inputenc}
\usepackage{graphicx}
\usepackage{float}
\usepackage{geometry}
\usepackage{parskip}
\usepackage{placeins}
\usepackage{braket}
\usepackage{amsmath}
\usepackage{amsfonts}
\usepackage{bm}
\usepackage{soul}
\usepackage{xcolor}
\usepackage{esvect}
\usepackage{tabularx}
\usepackage{makecell}

\usepackage{multirow}
\usepackage{outlines}
\usepackage{subcaption}
\usepackage[hidelinks]{hyperref}
\usepackage{cleveref}
\usepackage{authblk}
\usepackage{numprint}
\usepackage{comment}
\usepackage[makeroom]{cancel}
\usepackage{caption}
\usepackage{booktabs}       
\usepackage[backend=biber, sorting=none, style=numeric-comp, maxcitenames=1]{biblatex}
\usepackage{csquotes}
\Crefname{figure}{Fig.}{Figs.}
\Crefname{table}{Tab.}{Tabs.}
\Crefname{equation}{Eq.}{Eqs.}
\Crefname{section}{Sec.}{Secs.}

\crefname{section}{sec.}{secs.}
\crefname{figure}{fig.}{figs.}
\crefname{table}{tab.}{tabs.}
\crefname{equation}{eq.}{eqs.}

\def\ev#1{\mathinner{\langle{#1}\rangle}}
\def\Ev#1{\left\langle#1\right\rangle}

\npthousandsep{,}

\newcommand{\citeauthorname}[1]{{\citeauthor{#1}~\cite{#1}}}

\renewcommand{\vec}[1]{{\bm{#1}}}

\DeclareMathOperator*{\argmin}{argmin}

\newcommand{\nel}{{n_{\mathrm{el}}}}
\newcommand{\norb}{{n_{\mathrm{orb}}}}
\newcommand{\nbasis}{{n_{\mathrm{b}}}}
\newcommand{\nnuc}{{N_{\mathrm{atoms}}}}

\newcommand{\latsc}{{\vec{L}_\mathrm{sc}}}

\newcommand{\ndet}{n_{\mathrm{det}}}

\newcommand{\nemb}{{d_{\mathrm{emb}}}}
\newcommand{\norbfeat}{d_{\text{orb}}}

\newcommand{\rvec}{\vec{r}}
\newcommand{\Rvec}{\vec{R}}
\newcommand{\lat}{\vec{L}}

\newcommand{\dr}{\text{d}\vec{r}}
\newcommand{\ktwist}{\vec{k}_\mathrm{s}}
\renewcommand{\Re}{\operatorname{Re}}

\title{Transferable Neural Wavefunctions for Solids}
\author[$\dagger \ast$]{L. Gerard}
\author[$\dagger \ast$]{M. Scherbela}
\author[$\ddagger \ast$]{H. Sutterud}
\author[$\ddagger$]{W.M.C. Foulkes}
\author[$\dagger$, $\mathparagraph$]{P. Grohs}

\affil[$\dagger$]{Faculty of Mathematics, University of Vienna, Oskar-Morgenstern-Platz 1, A-1090 Vienna, Austria}
\affil[$\ddagger$]{Department of Physics, Imperial College London, South Kensington Campus, London SW7 2AZ}
\affil[$\mathparagraph$]{Johann Radon Institute for Computational and Applied Mathematics,  Austrian Academy of Sciences, Altenbergerstrasse 69, 4040 Linz, Austria}
\affil[$\ast$]{These authors contributed equally}

\date{}   
\begin{document}

\maketitle

\begin{abstract}
Deep-Learning-based Variational Monte Carlo (DL-VMC) has recently emerged as a highly accurate approach for finding approximate solutions to the many-electron Schrödinger equation.
Despite its favorable scaling with the number of electrons, $\mathcal{O}(\nel^{4})$, the practical value of DL-VMC is limited by the high cost of optimizing the neural network weights for every system studied.
To mitigate this problem, recent research has proposed optimizing a single neural network across multiple systems, reducing the cost per system.
Here we extend this approach to solids, where similar but distinct calculations using different geometries, boundary conditions, and supercell sizes are often required.
We show how to optimize a single ansatz across all of these variations, reducing the required number of optimization steps by an order of magnitude.
Furthermore, we exploit the transfer capabilities of a pre-trained network.
We successfully transfer a network, pre-trained on $2 \times 2 \times 2$ supercells of LiH, to $3 \times 3 \times 3$ supercells.
This reduces the number of optimization steps required to simulate the large system by a factor of 50 compared to previous work.
\end{abstract}

\section{Introduction}

Many interesting material properties, such as magnetism and superconductivity, depend on the material's electronic structure as given by the ground-state wavefunction.
The wavefunction may in principle be found by solving the time-independent Schrödinger equation, but doing so with sufficient accuracy is challenging because the computational cost grows dramatically with the number of particles. 
The challenge is particularly pronounced in solid state physics, where accurate calculations for periodic systems require the use of large supercells --- and consequently many particles --- to minimize finite-size effects.

Over the past few decades, density functional theory (DFT) has emerged as the primary workhorse of solid state physics. When using local or semi-local exchange-correlation functionals such as LDA or PBE \cite{perdewGeneralizedGradientApproximation1996a}, DFT calculations have a favorable scaling of $\mathcal{O}(\nel^3)$ or better, where $\nel$ is the number of electrons in the system, and an accuracy that is often sufficient to help guide and predict experiments \cite{sherrillFrontiersElectronicStructure2010,perdewGeneralizedGradientApproximation1996a}.
However, the choice of functional is in practice an uncontrolled approximation and DFT sometimes yields quantitatively or even qualitatively wrong results, especially for strongly correlated materials \cite{simonscollaborationonthemany-electronproblemGroundStatePropertiesHydrogen2020,burkePerspectiveDensityFunctional2012}.

Another approach, known as variational Monte Carlo (VMC), uses an explicit parameterized representation of the full many-body wavefunction and optimizes the parameters using the variational principle. This method has a favorable scaling of $\mathcal{O}(\nel^{3-4})$ \cite{foulkesQuantumMonteCarlo2001, pfauInitioSolutionManyelectron2020} but is limited in accuracy by the expressivity of the ansatz used.
Recently, deep neural networks have been employed as wavefunction ansätze \cite{carleo_science, pfauInitioSolutionManyelectron2020, hermannDeepneuralnetworkSolutionElectronic2020} and used to study a large variety of systems including small molecules \cite{pfauInitioSolutionManyelectron2020, gerardGoldstandardSolutionsSchrodinger2022, vonglehnSelfAttentionAnsatzAbinitio2022}, periodic model systems described by lattice Hamiltonians \cite{carleo_science, chooSymmetriesManyBodyExcitations2018, sharirDeepAutoregressiveModels2020, luoBackflowTransformationsNeural2019}, the homogeneous electron gas \cite{cassella_model_solids_physrevlett_2023, pescia2023heg}, and Fermi liquids \cite{kim2023neuralnetwork, lou2024neural}.
 Thanks to their flexibility and expressive power, deep-learning-based VMC (DL-VMC) approaches provide the best current estimates for the ground-state energies of several small molecules \cite{gerardGoldstandardSolutionsSchrodinger2022, vonglehnSelfAttentionAnsatzAbinitio2022}

Efforts to apply DL-VMC to real solids \cite{li_ab_solids_2022, liElectricPolarizationManyBody2023} have been limited by the high computational cost involved.
While a single calculation may be feasible, studying real solids requires many similar but distinct calculations. First, it is necessary to perform calculations involving increasingly larger supercells to estimate finite-size errors and extrapolate results to the thermodynamic limit. Second, twist-averaged boundary conditions (TABC) are used to accelerate the rate at which the finite-size errors reduce as the supercell size increases \cite{linTwistaveragedBoundaryConditions2001}. This requires averaging the results for each supercell over many calculations using different boundary conditions. Lastly, studying a given system often requires calculations for different geometries and lattice constants. 
Since most existing DL-VMC ansätze require optimizing a new wavefunction from scratch for each new system, the computational cost quickly becomes prohibitive even for systems of moderate size.
For example, \citeauthor{li_ab_solids_2022} proposed DeepSolid \cite{li_ab_solids_2022}, an ansatz that can accurately model periodic wavefunctions with up to 100 electrons, but required over 80k GPU hours to study a single system. 

In this work we implement a transferable DL-VMC ansatz for real solids that takes as input not only the electron positions but also other parameters of the system, such as its geometry or boundary condition. The key idea, based on \cite{scherbela2023foundation} and detailed in \Cref{sec:architecture}, is to map computationally cheap, uncorrelated mean-field orbitals to expressive neural network orbitals that depend on the positions of all of the electrons.
The transferability of the network orbitals allows us to optimize a single ansatz for many variations of unit-cell geometry, boundary condition and supercell size all at once. Because the ansatz learns to generalize across systems, we can use pre-trained models as highly effective initializers for new systems or larger supercells.

Compared to previous DL-VMC work without transferability, our approach yields more accurate results, gives access to denser twist averaging (reducing finite-size effects), and requires a fraction of the computational resources.
For example, for lithium hydride, transferring a 32 electron calculation to one with 108 electrons yields more accurate results than previous work \cite{li_ab_solids_2022} at $\approx 1/50$ of the computational cost.

This paper is structured as follows.
\Cref{sec:results} describes the results of applying the transferable DL-VMC ansatz to three different systems: 1D hydrogen chains, 2D graphene, and 3D lithium hydride.
In \Cref{sec:discussion} we discuss the implications of the results, limitations of the ansatz, and possible future work.
\Cref{sec:methods} explains the DL-VMC approach and our ansatz, as well as other technical details of our work.

\section{Results}
\label{sec:results}

\subsection{1D: Hydrogen chains}
\label{sec:hydrogen_chains}
Chains of hydrogen atoms with periodic boundary conditions provide a simple 1D toy system that nevertheless exhibits rich physics such as dimerization, a lattice-constant-dependent metal-insulator transition, and strong correlation effects. A collaborative effort \cite{simonscollaborationonthemany-electronproblemSolutionManyElectronProblem2017, simonscollaborationonthemany-electronproblemGroundStatePropertiesHydrogen2020} has obtained results for this system using a large variety of high-accuracy methods, providing a trustworthy benchmark.

\paragraph{Energy per atom}
The first test is to obtain the total energy per atom for a fixed atom spacing, $R=1.8 a_0$ (where $a_0$ is the Bohr radius), in the thermodynamic limit (TDL) attained as the number of atoms in the supercell tends to infinity.
To this end, we train two distinct models on periodic supercells with $N_\text{atoms} = 4,6, \dots, 22$.
The first model is trained at twist $k=0$ (the $\Gamma$ point) only. The second is trained using all twists from a $\Gamma$-centered four-point Monkhorst-Pack grid \cite{monkhorstSpecialPointsBrillouinzone1976}. The three inequivalent twists are $k=0, \frac{1}{4}, \frac{1}{2}$ in units of $2\pi / R$, and their weights are $w = 1, 2, 1$, respectively.
Once the model has been pre-trained on these relatively short chains, we fine-tune it on larger chains with $N_\text{atoms} = 32, 38$.
We use the extrapolation method described in \cite{simonscollaborationonthemany-electronproblemSolutionManyElectronProblem2017} to obtain the energy $E_\infty$ in the TDL. This entails fitting a polynomial of the form $E = E_\infty + E_1 N_\text{atoms}^{-1} + E_2 N_\text{atoms}^{-2}$.
Previous authors have extrapolated the energy using only chain lengths of the form $N_\text{atoms} = 4n+2, \; n\in \mathbb{N}$, which have filled electronic shells. We also report extrapolations using chain lengths $N_\text{atoms} = 4n$, which lead to partially filled shells.

\begin{figure}[htb]
    \centering
    \begin{subfigure}[b]{\textwidth}
        \centering
        \includegraphics[width=0.475\linewidth]{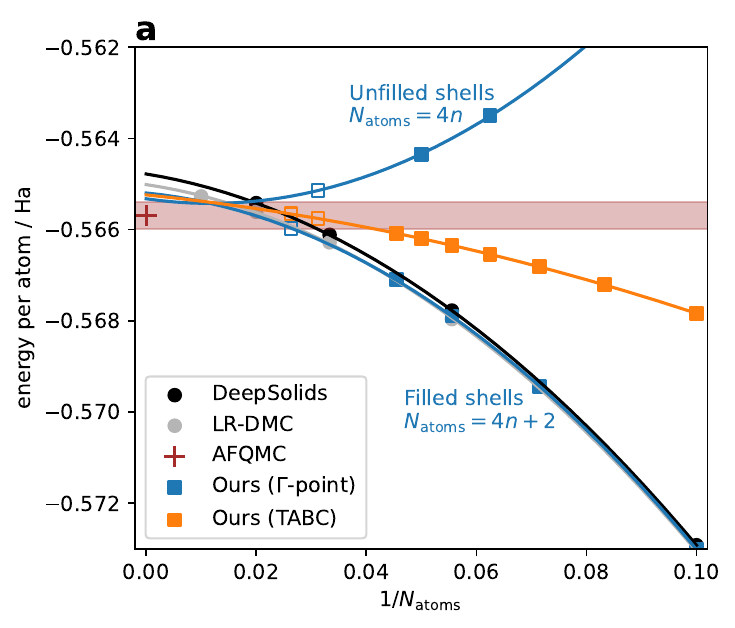}%
        \hfill
        \includegraphics[width=0.475\linewidth]{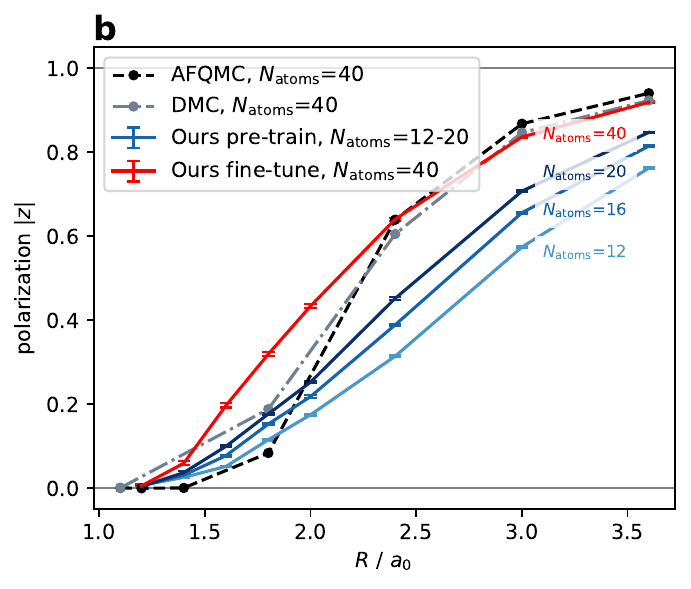}
    \end{subfigure}
    \caption{\textbf{1D Hydrogen chain}: \textbf{a}: Extrapolation of the energy per atom to the thermodynamic limit for $R=1.8 a_0$. Results obtained using DeepSolid (neural wavefunction), lattice-regularized diffusion Monte Carlo (LR-DMC), auxiliary field Monte Carlo (AFQMC), and our transferable neural wavefunction are shown. Open markers indicate energies computed by fine-tuning a model pre-trained on smaller supercells. The shaded area depicts the statistical uncertainty in the AFQMC result. Monte Carlo uncertainty of our results is $\approx10 \mu \text{Ha}$, well below the marker size.
    \textbf{b}: The complex polarization $|z|$ as a function of the inter-atomic separation, $R$, showing a phase transition between a metal at small $R$ and an insulator at large $R$. AFQMC, DMC, and VMC results are taken from \cite{simonscollaborationonthemany-electronproblemGroundStatePropertiesHydrogen2020}. DeepSolid results are taken from \cite{li_ab_solids_2022}.}
    \label{fig:hchains}
\end{figure}

\Cref{fig:hchains}a shows that all of our extrapolations ($\Gamma$-point filled shells, $\Gamma$-point unfilled shells, and TABC) are in good qualitative agreement with previous results obtained using methods such as lattice-regularized diffusion Monte Carlo (LR-DMC) \cite{simonscollaborationonthemany-electronproblemSolutionManyElectronProblem2017} and DeepSolid \cite{li_ab_solids_2022}, a FermiNet-based \cite{pfauInitioSolutionManyelectron2020} neural wavefunction for solids. 
Quantitatively, we achieve slightly lower (and thus, by the variational principle, more accurate) energies than DeepSolid for all values of $N_{\text{atoms}}$. Using TABC, we obtain $E_\infty = -565.24(2)\,\text{mHa}$, which is $0.2 - 0.5\,\text{mHa}$ lower than the estimate obtained using LR-DMC and DeepSolid, and agrees within uncertainty with the extrapolated energy computed using the auxiliary-field quantum Monte Carlo (AFQMC) method \cite{simonscollaborationonthemany-electronproblemSolutionManyElectronProblem2017}.
Most notably, though, we obtain these results at a fraction of the computational cost of DeepSolid. Whereas DeepSolid required a separate calculation with 100,000 optimization steps for each value of $N_\text{atoms}$ (and would have required even more calculations for twist-averaged energies), we obtain results for all 10 chain lengths and values of $N_\text{atoms} = 4,\dots,22$, with 3 twists for each system, using only 50,000 optimization steps in total.
Furthermore, by re-using the model pre-trained on smaller chains, we obtain results for the larger chains with $N_\text{atoms}=32,\,38$ using only 2,000 additional steps of fine tuning. This reduces the cost of simulating the large chains by a factor of approximately 50.
We note that, as expected, the use of TABC reduces the finite-size errors, allows to combine results for filled and unfilled shells in the extrapolation, and leads to faster and more uniform convergence of the energy per atom.

\paragraph{Metal-insulator transition}
The 1D hydrogen chain exhibits a transition from an insulating phase at large inter-atomic separation, $R$, to a metallic phase at small $R$. The transition can be quantified by evaluating the complex polarization along the length of the chain
\begin{equation} \label{eq:complex_polarization}
    z = \Braket{e^{i\frac{2\pi}{R N_\text{atoms}} \sum_{i=1}^\nel x_{i}}}, 
\end{equation}
where $x_i$ is the position of electron $i$ in the direction of the chain. 
The expectation value is defined as $\Braket{\dots} \equiv \int \Psi^*(\vec r) \dots \Psi(\vec r) \,\mathrm{d}\vec r$, where $\vec{r} = (\vec{r}_1, \vec{r}_2, \ldots, \vec{r}_{\nel})$ is a $3\nel$-dimensional vector of electron positions, $\Psi$ is the (approximate) ground-state wavefunction, and the integral is over all $3\nel$ electronic degrees of freedom.
Although the polarization is easy to evaluate in principle, studying the transition is computationally costly because it requires many similar but distinct calculations: multiple values of $R$ are required to locate the transition; multiple twists $k$ are required to obtain accurate twist-averaged polarizations; and multiple chain lengths $N_{\text{atoms}}$ are required to allow extrapolation to the TDL.
Even for a modest selection of all of these variations, studying the phase transition in detail requires hundreds of calculations.
Using our transferable wavefunction, on the other hand, allows us to train a single model to represent the wavefunction for all parameter variations at once.

We trained a single ansatz to describe all 120 combinations of: (a) 3 distinct chain lengths, $N_{\text{atoms}}=12,16,20$; (b) 5 symmetry-reduced $k$-points of an 8-point $\Gamma$-centered Monkhorst-Pack grid; and (c) 8 distinct atom spacings between $R=1.2 a_0$ and $R=3.6 a_0$.
A total of 200k optimization steps were carried out, after which the complex polarization was evaluated using \Cref{eq:complex_polarization}.
To improve our estimates for $N_{\text{atoms}}\to\infty$, we fine-tuned this pre-trained model for 2k steps on chain lengths of $N_\text{atoms}=40$ and a denser 20-point Monkhorst-Pack grid containing 11 symmetry-reduced twists.
\Cref{fig:hchains}b shows that our approach qualitatively reproduces the results obtained using DMC and AFQMC. In agreement with Motta et al.~\cite{simonscollaborationonthemany-electronproblemGroundStatePropertiesHydrogen2020}, we observe a second-order metal-insulator transition. However, where Motta estimates the critical atom spacing $R_\text{crit}=1.70(5)a_0$, our results are more consistent with $R_\text{crit}=1.35(5)a_0$.
A possible explanation for the disagreement is that our neural wave function may be less accurate (and may therefore produce relatively higher energies) for metals than insulators, disfavoring the metallic phase.
Another possible explanation follows from the observation that, unlike the VMC method used here, the DMC and AFQMC methods yield biased estimates of the expectation values of operators, such as the complex polarization, that do not commute with the Hamiltonian \cite{foulkesQuantumMonteCarlo2001, shiAFQMC2021}.

Also in agreement with Motta et al.~\cite{simonscollaborationonthemany-electronproblemGroundStatePropertiesHydrogen2020}, we find that the hydrogen chain has an antiferromagnetic ground state at large lattice constant $R$. 
The expected atomic spins are zero on every atom, but the spins on neighbouring atoms are antiferromagnetically correlated.
As the lattice constant gets smaller and the system transitions to the metallic phase, these correlations decrease.

\subsection{Graphene}
\label{sec:graphene}
To demonstrate the application of our transferable DL-VMC ansatz to a two-dimensional solid, we compute the cohesive energy of graphene in a $2 \times 2$ supercell and compare against the DL-VMC results of Li et al.~\cite{li_ab_solids_2022}. 
We employ twist-averaged boundary conditions, apply structure-factor-based finite-size corrections \cite{chiesaFiniteSizeErrorManyBody2006} as detailed in the methods section, and add zero-point vibrational energies (ZPVE) (see \Cref{sub:zpve}). 
Li et al.\ restricted their calculation to a Monkhorst-Pack grid of $3 \times 3$ twists, yielding three symmetry-reduced twists in total. 
In our case, since we are able to compute multiple twists at once with minimal extra cost, we increase the grid size to $12 \times 12$. 
This increases the number of symmetry-reduced twists approximatly 6 times, from 3 to 19. The larger twist grid contains a subset of the twists considered by Li et al., allowing a direct comparison with their independent energy calculations. 

In \Cref{tab:graphene_tabc_results} (see also \Cref{sec:si_total_energies} for separate twist energies without ZPVE), we compare total energies calculated at each of the three twists on the $3 \times 3$ twist grid and cohesive energies obtained by averaging over the $3 \times 3$ and $12 \times 12$ twist grids.
We find that our energies for $k_1$ and $k_2$ are lower than the energies obtained by DeepSolid by 1~mHa and 7 mHa / primitive cell respectively, while our energy for $k_3$ is higher by 4~mHa. 
Overall this leads to a twist-averaged energy which is 4~mHa / primitive cell lower than the DeepSolid energy.
This is a direct consequence of how our approach divides the total number of optimization steps across twists. 
At every optimization step we randomly sample a symmetry-reduced twist to optimize next. 
The sampling probability is obtained by normalizing the weights assigned to the twists on the symmetry-reduced $12 \times 12$ Monkhorst-Pack grid.
This procedure ensures that more optimization steps are spent on twists with high contribution to the final energy and fewer steps on less important twists.
The $k_3$ twist is chosen in around $\sim$2.0\% of all optimization steps, whereas the second twist is chosen two times more often. 
The wave function at $k_2$ is therefore optimized more stringently. 

Furthermore, we obtain energies not only for the $3\times3$ Monkhorst-Pack-grid, but also for the full $12\times12$-grid, allowing us to assert convergence with respect to the k-point density.
We stress that we only required a single neural network, optimized for 120k steps to obtain energies for all twists (both the $12\times12$-grid and the $3\times3$-subset).
DeepSolid on the other hand optimized for 900k steps in total, obtaining energies only for the $3\times3$-twist-grid.

We compare against experimental cohesive energies obtained from thermochemistry data for graphite \cite{brewerCohesiveEnergiesLBL3720} corrected for the small inter-layer binding energy of $3.5$~mHa obtained using the Random Phase Approximation \cite{lebegueCohesivePropertiesAsymptotics2010}.
When computing cohesive energies and correcting for finite-size effects using a structure-factor-based correction and ZPVE (see \Cref{sub:sfc}) we obtain energies that are $7$~mHa lower than experimental values, i.e. we predict slightly stronger binding than experiment.
We hypothesize that this small discrepancy may be a finite size artifact stemming from the relatively small 2x2 supercell.

\begin{table}[htb]
    \centering
    \caption{
    \textbf{Total and cohesive energies of graphene} in Hartrees. The upper block of the table compares our results against the total energies computed by Li et al.~\cite{li_ab_solids_2022} at the three symmetry-inequivalent twists on the $3 \times 3$ Monkhorst-Pack grid. The twists are expressed in the basis of the reciprocal lattice vectors.
    The lower block compares the twist-averaged cohesive energy per primitive cell with experimental results, showing the effect of increasing the size of the twist grid from $3 \times 3$ to $12 \times 12$. For the calculation of the cohesive energy we follow \citeauthorname{li_ab_solids_2022} and take as the energy of a single carbon atom E$=-37.84471$ Ha \cite{pfauInitioSolutionManyelectron2020}. All results include a structure-factor-based finite-size correction \cite{chiesaFiniteSizeErrorManyBody2006} and ZPVE (see \Cref{sub:sfc}). The experimental results are based on \cite{brewerCohesiveEnergiesLBL3720, lebegueCohesivePropertiesAsymptotics2010}.}
    \begin{tabular}{llrrrr}
    & \textbf{Twists} & \textbf{Weight} &\textbf{DeepSolid} & \textbf{Our work} & \textbf{Exp.}\\ 
    \toprule
    \multirow{3}{*}{Total energy / Ha} & $\vec{k}_1=(0, \ 0)$ & $1/9$ & $-76.1406$  & 
    $-76.1414(2)$ &  -\\ 
     & $\vec{k}_2=(1/3, 1/3)$ & $2/3$ & $-76.2342$ & $-76.2415(2)$ & -\\
     & $\vec{k}_3=(2/3, 1/3)$ & $2/9$ & $-76.2479$ & $-76.2432(2)$ & -\\ \midrule
    \multirow{2}{*}{Cohesive energy / Ha} & Averaged $3 \times 3$ & - & $-0.5375$  & $-0.5413(2)$ & \multirow{2}{*}{$-0.538(2)$} \\ 
     & Averaged $12 \times 12$& - & - & $-0.5451(2)$ & \\ 
    \midrule
    \bottomrule
    \label{tab:graphene_tabc_results}
    \end{tabular}
\end{table}

\begin{figure}[htb]
    \centering
    \includegraphics[width=1.0\linewidth]{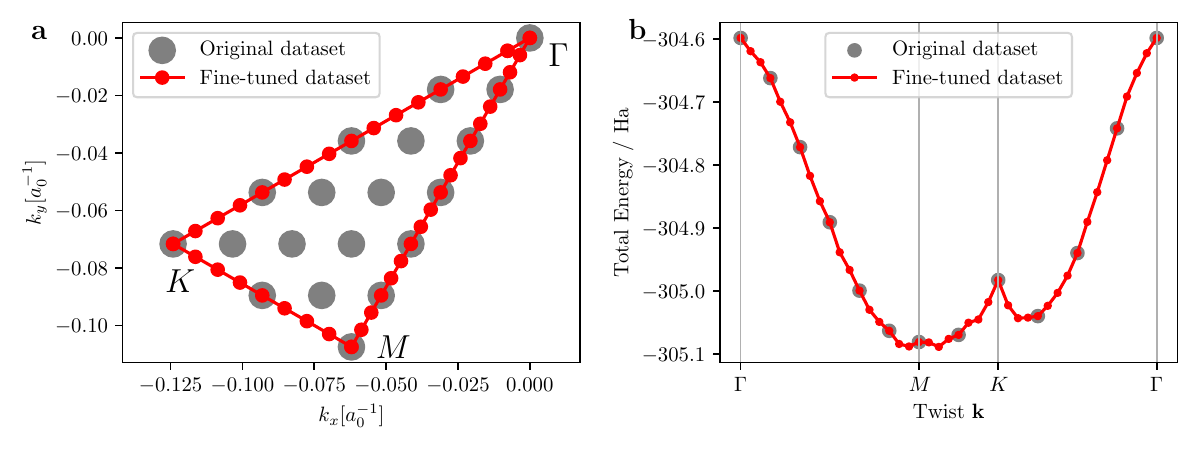}
    \caption{\textbf{Twist-dependent energy of Graphene} \textbf{a:} Grid of pretrained twists and path of fine-tuned values through Brillouin zone. \textbf{b:} Fine-tuned energies of graphene along path of twists across the Brillouin zone. Fine-tuned using shared optimization and around 100 additional optimization iterations per twist. Error bars are smaller than the size of the markers. }
    \label{fig:banddiagram}
\end{figure}

With a network that has been trained across the entire Brillouin zone, we can  evaluate observables along arbitrary paths in $k$ space. \Cref{fig:banddiagram} is a bandstructure-like diagram, showing how the total energy varies along a path passing through the high-symmetry $k$ points $\Gamma = (0, 0)$, $M= (0, 1/2)$, and $K = (1 / 3, 2 / 3)$ in units of the supercell reciprocal lattice vectors.
We use the pre-trained model from the 12x12 Monkhorst-Pack grid and transfer it to the bandstructure-like diagram with k-points previously unseen during optimization, requiring only a few additional optimization steps. 
We fine-tune the pre-trained model for the k-points on the path, using around 100 optimization steps per twist and then evaluate the energies along the path.
Analogously to the Dirac cone visible in the one-electron bandstructure, also our many-electron bandstructure displays a characteristic cusp at the $K$ point.
However, since we are plotting the dependence of an $\nel$-electron energy on a many-body parameter (the twist), \Cref{fig:banddiagram} is not directly comparable to a conventional one-electron bandstructure diagram.

\FloatBarrier
\subsection{Lithium Hydride}
\label{sec:lih}
\begin{figure}[!htpb]
    \centering
    \includegraphics[width=0.5\textwidth]{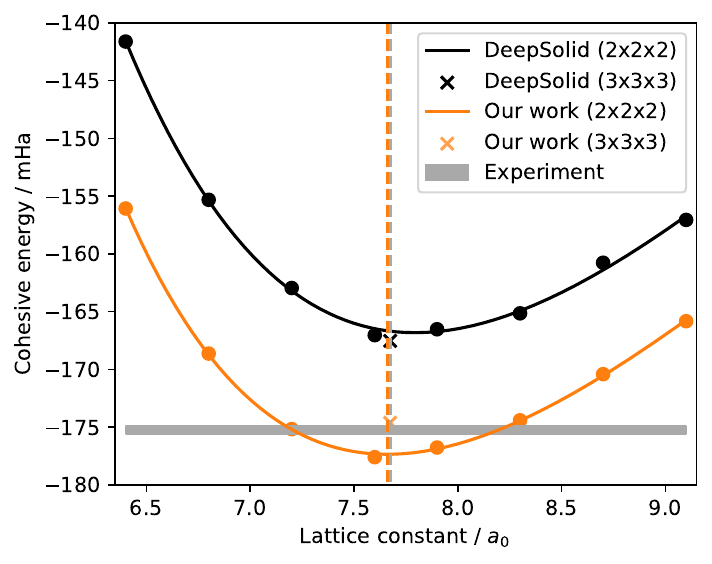}
    \caption{\textbf{Energy-volume curve of LiH per primitive cell} for a $2 \times 2 \times 2$ supercell as calculated using DeepSolid \cite{li_ab_solids_2022} and our transferable DL-VMC method. The DeepSolid results (black circles, with a Birch-Murnaghan fit represented as a black line) were obtained at a single twist, the $\Gamma$-point. Hartree-Fock corrections were applied, as discussed in \cite{li_ab_solids_2022}, and a ZPVE correction added. Our results (orange circles, with a Birch-Murnaghan fit represented as an orange line) are twist averaged, using a $5 \times 5 \times 5$ Monkhorst-Pack grid per lattice constant. Structure-factor-based corrections were applied and a ZPVE correction added. The grey bar indicates the experimental uncertainty \cite{Nolan_exp_results_lih}.  The statistical error bars are too small to be visible on this scale and therefore have beenn omitted. The vertical orange dashed line indicates the equilibrium lattice constant as calculated from the Birch-Murnaghan fit to our data. The orange cross shows the twist-averaged cohesive energy of a $3 \times 3 \times 3$ simulation cell, using again structure factor correction. This was obtained by transferring the network pre-trained for the $2 \times 2 \times 2$ system to a $3 \times 3 \times 3$ supercell, using only 8,000 additional optimization steps.  A $5 \times 5 \times 5$ Monkhorst-Pack grid of twists was used. The black cross shows the result of DeepSolid's $3 \times 3 \times 3$ $\Gamma$-point calculation with a Hartree-Fock finite-size correction.}
    \label{fig:lih_twist_avg_pes}
\end{figure}

We have also used the transferable DL-VMC ansatz to evaluate the energy-volume curve of LiH in the rock-salt crystal structure. As shown in \Cref{fig:lih_twist_avg_pes} (see also \Cref{sec:si_total_energies}), we obtain the energy-volume curve by fitting a Birch-Murhaghan equation of state to the total energies of a $2 \times 2 \times 2$ supercell at eight different lattice parameters.
To reduce finite-size errors, the eight total energies are twist averaged using a $5 \times 5 \times 5$ $\Gamma$-centered Monkhorst-Pack grid and include structure-factor-based finite-size corrections. 
For comparison, \citeauthor{li_ab_solids_2022} \cite{li_ab_solids_2022} performed a $\Gamma$-point calculation only and estimated finite-size errors by converging a Hartree-Fock calculation with an increasingly dense twist grid. 
To all results we add zero-point vibrational energies (ZPVE) taken from \cite{Nolan_exp_results_lih}, making the calculated cohesive energy less negative by $\approx 8$ mHa. 
The work of \citeauthor{li_ab_solids_2022} \cite{li_ab_solids_2022} took no account of the ZPVE, explaining the slight difference between our depiction of their results, shown in \Cref{fig:lih_twist_avg_pes}, and their original publication \cite{li_ab_solids_2022}.

We trained a single neural network wavefunction across 8 lattice constants and 10 symmetry-reduced twists, making 80 systems in total. 
In comparison, \citeauthor{li_ab_solids_2022} \cite{li_ab_solids_2022} required a separate calculation for each geometry.

The Birch-Murnaghan fit gives an equilibrium lattice constant of $7.66(1) a_0$ (dotted orange line), which agrees well with the experimental value of $7.674(2) a_0$ \cite{Nolan_exp_results_lih}. 
Our Birch-Murnaghan estimate of the cohesive energy of $-177.3(1)$~mHa / primitive cell deviates from the experimental value of $-175.3(4)$~mHa  by $-2.0(5)$~mHa.
This marks an improvement over the DeepSolid results \cite{li_ab_solids_2022} of $-166.8(1)$ mHa, which differ from experiment by $8.5(5)$ mHa. 
Because we are able to optimize all systems at once, our results were obtained with roughly 5\% of the compute required by DeepSolid.

Although we improve on the DeepSolid baseline, the cohesive energy might potentially still be impacted by finite-size effects because of the small size of the $2 \times 2 \times 2$ supercell used. To check this, we also studied a larger supercell containing $3 \times 3 \times 3$ primitive unit cells.

Previous work on molecules has shown that it is sometimes possible to transfer pre-trained neural wavefunctions to larger systems. 
For example, transferring parameters from a wavefunction pre-trained on small molecules to larger molecules allowed a reduction in the number of optimization steps by an order of magnitude compared to a random initialization of parameters \cite{scherbela2023variationalOnABudget, scherbela2023foundation}. 
For our application, we transferred the neural wavefunction optimized to represent a $2 \times 2 \times 2$ simulation cell of LiH at multiple twists and lattice constants to a much larger $3 \times 3 \times 3$ supercell.
The 108-electron $3 \times 3 \times 3$ system is one of the largest to have been studied using neural wavefunctions to date, with over three times more electrons than the 32-electron $2 \times 2 \times 2$ system used for pre-training. 
DeepSolid used $400,000$ optimization steps to get an estimate for the cohesive energy and overestimated the energy by around $7$ mHa / per primitive cell compared to the experimental results \cite{Nolan_exp_results_lih, li_ab_solids_2022}.
By contrast, starting with the converged neural wavefunction for the $2 \times 2 \times 2$ supercell, we are able to calculate the cohesive energy for the $3\times3\times3$-supercell with only $8,000$ additional optimization steps shared across ten different twists.
Using twist averaging, a structure-factor correction, and a ZPVE correction as before, we obtain a cohesive energy of $-$174.6~mHa / primitive cell, deviating from experiment by only 0.7(5)~mHa / primitive cell. 
The magnitude of this deviation is close to the 0.4 mHa spread of experimental data obtained from different thermochemistry experiments \cite{Nolan_exp_results_lih}.
Our twist-averaged $3 \times 3 \times 3$ calculation required only $\sim$2\% of the computational resources used by \citeauthor{li_ab_solids_2022} \cite{li_ab_solids_2022} for a single $\Gamma$-point calculation.

\section{Discussion}
\label{sec:discussion}

Previous DL-VMC approaches were only capable of computing the wavefunction of a single system at once \cite{li_ab_solids_2022} or were limited to gas-phase molecules \cite{scherbela2023variationalOnABudget, scherbela2023foundation}. In this work we propose a generalized neural network-based wavefunction for periodic systems. In addition to the electron positions, the network uses information about the Hartree-Fock one-electron eigenfunctions. Maximally localized Wannier functions are generated from the occupied Hartree-Fock orbitals and represented as expansions in a basis of local atomic-like orbitals. The expansion coefficients are then used as network inputs.
By mapping cheap and low-accuracy orbital descriptors to highly accurate deep-learning-based orbitals, in a manner based on the ideas of Ref.~\cite{scherbela2023foundation}, we are able to optimize a single neural wavefunction model across twists, lattice constants and supercell sizes all at once.
This reduces the computational cost by an order of magnitude. Cost reductions are particularly important in solid-state simulations because large supercells must be studied to minimize finite-size effects.

We also investigate the transfer capabilities of a pre-trained wavefunction model. In particular, we transfer a pre-trained model to a system more than three times larger and find that energies converge using more than 50 times fewer optimization steps.
This could pave the way to simulations of the large supercells required to study  metals or perhaps even high-temperature semiconductors.

Furthermore, we find that our approach is able to represent qualitatively different wavefunctions within a single model. For example, for the hydrogen chain, a single ansatz with identical parameters can represent both the metallic state and the insulating state. A concurrent pre-print \cite{rende2024finetuning} exploring a related idea --- pre-training a neural wavefunction embedding layer for a lattice model near a phase transition and then fine-tuning a final layer on either side of the transition --- may shed some light on this success. Our embeddings can learn robust, transferable, features that allow efficient representation of the wavefunction in either phase.

Besides the transfer to larger systems, our method allows for efficient fine-grained twist-averaged calculations. Unlike previous DL-VMC methods, we use a single network to represent wavefunctions at multiple different twists, avoiding the computational overhead of training separate neural network-based wavefunctions at each twist.
Our ansatz already allows the concurrent optimization of systems with varying numbers of particles, so this approach could be extended to grand-canonical twist averaging \cite{azadiEfficientMethodGrandcanonical2019}, in which the number of electrons in the supercell varies with the twist.

Our approach shares many of the limitations of other DL-VMC methods, including the sensitivity with regard to MCMC initialization. A standard practice in DL-VMC is to assign each electron a spin and initialize it close to the nuclei at the beginning of the calculation. If the electrons are initialized in an anti-ferromagnetic pattern, i.e., alternating the spins of neighboring atoms, but the ground state is ferromagnetic, as can be the case for the hydrogen chain when the inter-atomic separation is small (see \Cref{sec:hydrogen_chains}), our approach tends to converge to local minima. FermiNet suffers from similar problems.

Another limitation arises from the allocation of compute budget between the multiple geometries or systems described by a single neural network. As discussed in \Cref{sec:graphene}, we allocate more compute during optimization to twists with a larger weight. This has a positive effect on twist-averaged results in general, because twists with higher contribution are converged to higher accuracy (see \Cref{tab:graphene_tabc_results}). 
However for individual twists, when plotting for example the band structure (see \Cref{fig:banddiagram}), not all twists are optimized to same accuracy potentially skewing results. 

In comparison with other DL-VMC methods, the construction of our orbital matrix introduces an additional dependency on the positions of the nuclei (see \Cref{eq:orbitals_tao}). Tests suggest that this worsens the empirical scaling of computational cost with the number of particles compared to FermiNet, and that the added cost is greatest for systems with a low ratio of electrons to nuclei. 
In the Supplementary Material, S1, we investigate empirically the effect on scaling of our newly proposed architecture. The overall scaling is in principle still dominated by the computation of the determinant and its derivative, implying that the time per iteration is $\mathcal{O}(\nel^4)$, just as it is for FermiNet. For the system sizes investigated, however, we observe that our method requires approximately twice (the ratio is system dependent) as much run time per optimization step as FermiNet. This higher per-iteration cost is fortunately small relative to the orders of magnitude reduction in the total number of iterations steps required.

In summary, the approach introduced in this paper reduces the computational cost of DL-VMC simulations and allows them to be used to study larger supercells with more twists. 
By combining the present work with the efficient forward evaluation of the Laplacian of the wavefunction recently introduced by Li et al.~\cite{li2023forward} and using pseudo-potentials \cite{li_effective_core_potential} to represent the core electrons, it should be possible to scale to even larger systems. 

\section{Methods}
\label{sec:methods}
\sloppy
\subsection{Notation}
All vectors, matrices and tensors are denoted by \textbf{bold letters}, except for functions. 
We use lower-case indices $i, j = 1, \dots, \nel$ for electron positions and upper case indices $I, J=1, \dots, \nnuc$ for atom positions, where $\nel$ and $\nnuc$ are the numbers of electrons and atoms in the supercell. 
Orbitals are enumerated by the indices $\mu$ and $\nu$, which range from $1$ to $\nel$. 
The position of the $i$'th electron is $\vec{r}_i \in \mathbb{R}^3$. When $i$ is not used as a subscript it denotes the imaginary unit. 
By $\vec{r} = (\vec{r}_1, \dots, \vec{r}_{\nel})$ we denote the $3n_{\text{el}}$-dimensional vector of all electron positions. 
Similarly, nuclear positions and charges are represented by $\vec{R} = (\vec{R}_1, \dots, \vec{R}_{\nnuc})$ and $\vec{Z} = (Z_1, \dots, Z_{\nnuc})$. 
The matrix $\vec{L} \in \mathbb{R}^{3 \times 3}$ contains the supercell lattice vectors in its columns. 
The twist vector, which may always be reduced into the first Brillouin zone of the supercell, is denoted by $\ktwist$. The dot product of two vectors $\vec{a}$ and $\vec{b}$ is written $\vec{a}\cdot\vec{b}$ and by $\odot$ we refer to the element-wise multiplication (Hadamard product).

\subsection{Deep-learning Variational Monte Carlo}
The time-independent Schrödinger equation for a solid takes the form
\begin{align}
    \hat{H} \Psi = E \Psi, \qquad \hat{H} = - \frac{1}{2}\sum_{i} \nabla^2_{\vec{r}_i} + \hat{V}_{\text{Coulomb}}
\end{align}
with the Hamiltonian in the Born-Oppenheimer approximation 
and Coulomb potential $ \hat{V}_{\text{Coulomb}}$. A finite supercell is used to approximate the bulk solid, and the Coulomb potential is evaluated using the Ewald method, as described in \cite{ewald1921, cassella_model_solids_physrevlett_2023}.

In this work we are interested in finding the lowest eigenvalue of the Schrödinger equation --- the ground-state energy, $E_0$ --- and the corresponding energy eigenfunction. 
To find an approximate solution, one can reformulate the Schr\"{o}dinger equation as a minimization problem using the Rayleigh-Ritz variational principle. Given an arbitrary anti-symmetric trial wavefunction, $\Psi_{\boldsymbol{\theta}}$, with $\boldsymbol{\theta}$ denoting, for example, the trainable parameters of a neural network, the best attainable approximation to the ground state may be found by minimizing the energy expectation value
\begin{align}
    \label{eq:rayleigh_ritz}
    L(\boldsymbol{\theta}) = \mathbb{E}_{\vec{r} \sim \Psi_{\boldsymbol{\theta}}^2} \biggl[\frac{\hat{H}\Psi_{\boldsymbol{\theta}}}{\Psi_{\boldsymbol{\theta}}} \biggr] \geq E_0
\end{align}
with respect to $\vec{\theta}$. 
An important constraint for the construction of the trial wavefunction arises from the Pauli exclusion principle, which states that the wavefunction must be antisymmetric with respect to the permutations of different electron coordinates \cite{pfauInitioSolutionManyelectron2020}. In \Cref{sec:architecture} we outline in detail the underlying architecture of our neural network-based wavefunction. As in previous work, we approximate the expectation value in \Cref{eq:rayleigh_ritz} using Monte Carlo integration with samples drawn from the $3n_{\text{el}}$-dimensional probability density $|\Psi_{\boldsymbol{\theta}}(\vec{r})|^2$ \cite{pfauInitioSolutionManyelectron2020, hermannDeepneuralnetworkSolutionElectronic2020}.

A list of all relevant hyperparameters can be found in the supplementary information S3. 
\subsection{Architecture}
\label{sec:architecture}

\paragraph{Overview}
Our ansatz can be broken down into the computation of periodic input features, the computation of embeddings $\vec{e}_{iJ}$ for each electron-nucleus pair, the computation of correlated orbitals, and the assembly of the final wavefunction $\Psi_{\theta}$ as a sum of Slater determinants. 
Each step serves a distinct purpose.

The input features enforce the periodic boundary conditions of the supercell.
To capture correlation effects, we use a neural network to map single-electron coordinates to vectors in a latent space.
These vectors, also known as embeddings, depend on the positions of all of the other electrons in a permutation equivariant way. Each embedding therefore contains information about the corresponding electron as well as its environment.
The embeddings are subsequently mapped to many-electron orbitals as outlined below.

\paragraph{Ansatz}

Our wavefunction ansatz is a sum of Slater determinants multiplied by a Jastrow factor,
\begin{align}
  \Psi(\vec{r}, \vec{R}, \vec{Z}, \ktwist) = e^{J(\vec{r})} \sum_{d=1}^{\ndet} \det{\vec{\Phi}_{d}(\vec r, \vec{R}, \vec{Z}, \ktwist)}.
\end{align}
The optimization is free to adjust the relative normalizations of the determinants in the unweighted sum, making it equivalent to a weighted sum of normalized determinants, as might be used in a configuration-interaction expansion.
The Jastrow factor $e^{J(\vec{r})}$ is node-less and follows the work of \citeauthor{hermannDeepneuralnetworkSolutionElectronic2020} \cite{hermannDeepneuralnetworkSolutionElectronic2020}, while the determinant enforces the fermionic antisymmetry.
Instead of using single-particle orbitals in the determinant, as in most quantum chemical approaches, we follow other neural wavefunction methods \cite{pfauInitioSolutionManyelectron2020} and promote every entry $\Phi_{d,i \mu}$ in the orbital matrix $\vec \Phi_d$ from a one-electron orbital, $\phi_{d, \mu}(\vec{r}_i)$, to a many-electron orbital, $\Phi_{d,i \mu}(\vec{r})$ (temporarily dropping the dependency on $\vec R$, $\vec Z$, and $\vec k_s$ for the sake of brevity).
The many-electron orbitals are permutation equivariant, such that applying a permutation $\pi$ to the electron position vectors permutes the rows of $\vec{\Phi}_{d}$ by $\pi$, i.e., $\Phi_{d,i \mu}(\vec r_{\pi(1)}, \dots, \vec r_{\pi(\nel)}) = \Phi_{d,\pi(i) \mu}(\vec r_1, \dots, \vec r_\nel)$. This ensures that the determinant has the correct fermionic symmetry.
Each entry is constructed as a linear combination of atom-centered functions with permutation equivariant dependencies on both electrons and atoms
\begin{align}
    \Phi_{d,i\mu}(\vec r, \vec{R}, \vec{Z}, \ktwist)
    &= e^{i  \ktwist \cdot \vec{r}_i } \sum_{J=1}^{\nnuc} \varphi_{d\mu}(\vec r_i, \{\vec r\}, \vec R_J, \{\vec R\}).
    \label{eq:orbitals_tao0}
\end{align}
Here, $\{\vec r\}$ and $\{\vec R\}$ denotes the (permutation invariant) set of electron and atom positions, respectively.
The phase factor enforces the twisted boundary conditions, as explained in \Cref{sub:tabc}.
To construct the $\varphi_{d\mu iJ} \equiv \varphi_{d\mu}(\vec r_i, \{\vec r\}, \vec R_J, \{\vec R\})$ using a neural network, we use an adaptation of the recently proposed transferable atomic orbital ansatz \cite{scherbela2023foundation, scherbela2023variationalOnABudget}.
The orbitals are written as the inner product of an electron-nuclear embedding $\vec{e}_{iJ} \in \mathbb {R}^{n_{\text{emb}}}$ and an orbital embedding $\vec W_{d \mu J} \in \mathbb{C}^{n_{\text{emb}}}$, multiplied by an exponential envelope,
\begin{align}
    \varphi_{d\mu iJ} = (\vec{W}_{d \mu J} \cdot \vec{e}_{iJ})  e^{-a_{d\mu J} ||\vec{s}_{iJ}||^{\text{per}}}.  \label{eq:orbitals_tao}
\end{align}
where $a_{d \mu J}$ is a learnable decay rate, $\vec{s}_{iJ}$ is the vector from nucleus $J$ to electron $i$, expressed in the basis of the supercell lattice vectors, and $||\vec{s}_{iJ}||^{\text{per}}$ is the modulus of $\vec{s}_{iJ}$ in a periodic norm explained below.
Both the orbital embedding $\vec{W}_{d \mu J}$ and the decay length $a_{d\mu J}$ depend on the orbital $\mu$ and atom $J$ and are different for each determinant $d$.

To obtain $\vec{W}_{d\mu J}$ and $a_{d \mu J}$ in a transferable way, we do not parameterize them directly but represent them as functions of some orbital-specific descriptor $\tilde{\vec{c}}_{\mu J} \in \mathbb{R}^{\norbfeat}$:
\begin{align}
  \vec{W}_{d\mu J} &= f^{\text{W}}_d\left(\tilde{\vec{c}}_{\mu J}\right),  & a_{d\mu J} &= f^{\text{a}}_d\left(\tilde{\vec{c}}_{\mu J}\right),
\end{align}
with $f^{\text{W}}: \mathbb{R}^{\norbfeat} \to \mathbb{C}^{\ndet\times\nemb}$ and $f^{\text{a}}: \mathbb{R}^{\norbfeat} \to \mathbb{R}^{\ndet}$ denoting simple multi-layer perceptrons.
The orbital embedding includes information about single-particle orbitals of the system calculated with a mean-field method, which is key for the transferability of the ansatz.
The inputs are the orbital features $\tilde{\vec{c}}_{\mu J} \in \mathbb{R}^{\norbfeat}$, which are concatenations of the expansion coefficients of the localized mean-field orbitals in an atom-centred basis set, the twist $\ktwist$, the mean position of orbital $\mu $, and the position of atom $J$, with a combined dimensionality of $\norbfeat$.
While all parameters and intermediate computations of our network are real-valued, the last layer of $f^{\text{W}}$ is complex-valued to allow the network to represent complex-valued wavefunctions.

An important difference with respect to previous neural network-based wavefunctions is the use of electron-nuclear embeddings $\vec{e}_{iJ}$, which describe the interaction between electron $i$ and nucleus $J$.
Other architectures such FermiNet, but also the more closely related transferable atomic orbital ansatz \cite{scherbela2023foundation}, use embeddings to represent the interactions of a single electron $i$ with all nuclei instead. 
However, when the embeddings are both invariant under permutation of nuclei (which we require for efficient transferability) and invariant under translation of particles by a supercell lattice vector (which we require to enforce boundary conditions), they become periodic on the primitive lattice (see the supplementary information S2), not just the supercell lattice.
This is too restrictive to represent correlation beyond a single primitive cell.
We therefore opt to use electron-nucleus embeddings that are equivariant under permutation of nuclei at some additional computational cost explained in \Cref{sec:limited_expressiveness}.

\paragraph{Input} \label{sec:abc} We require our representation of the difference vectors $\vec{r}_{ij} = \vec{r}_i - \vec{r}_j$, $\vec{r}_{iI} = \vec{r}_i - \vec{R}_I$ and $\vec{r}_{IJ} = \vec{r}_I - \vec{R}_J$ to be periodic with respect to the supercell lattice. This is accomplished using the approach introduced by
\citeauthorname{cassella_model_solids_physrevlett_2023}. The first step is to transform the coordinates into supercell fractional coordinates with $\vec{s}_{ij}= \lat^{-1} \vec{r}_{ij}$, $\vec{s}_{iI}= \lat^{-1} \vec{r}_{iI}$ and $\vec{s}_{IJ}= \lat^{-1} \vec{r}_{IJ}$. Periodic versions of the difference vectors are then obtained by applying sine and cosine element-wise,
\begin{align}
    &\omega(\vec{s}) := [\sin(2\pi \vec{s}), \cos(2 \pi \vec{s})] , \qquad \omega: \mathbb{R}^3 \to \mathbb{R}^6, \\
    &\vec{x}_{ij} := \omega(\vec{s}_{ij}), \quad \vec{x}_{iJ} := \omega(\vec{s}_{iJ}), \quad \vec{x}_{IJ} := \omega(\vec{s}_{IJ}),
\end{align}
whereas square brackets denotes the concatenation operator. For the distance, we use the following periodic norm
\begin{align}
    \left ( ||\vec{s}||^{\text{per}} \right )^2 = \sum_{l,p=1}^{3} \biggl ( \bigl ( 1 - \cos(2\pi s_l) \bigr) A_{lp} \bigl ( 1 - \cos(2\pi s_p) \bigr) + \sin(2\pi s_l) A_{lp} \sin(2\pi s_p) \biggl )
\end{align}
for a vector $\vec{s} \in \mathbb{R}^3$ with the lattice metric $\vec{A} := L L^T$. This norm is used to define the periodic distance features: 
\begin{align}
    x_{ij} = ||\vec{s}_{ij}||^{\text{per}}, \quad x_{iJ} = ||\vec{s}_{iJ}||^{\text{per}}, \quad x_{IJ} = ||\vec{s}_{IJ}||^{\text{per}}.
\end{align}

\paragraph{Embedding} The periodic input features are used to generate high-dimensional embeddings $\vec{e}_{iJ}$ for the construction of the orbital matrix. The following embedding is a slight adaption of the appraoch used in the recently proposed Moon architecture \cite{gaoGeneralizingNeuralWave2023}. We start by aggregating the electron-electron features into message vectors $\vec{m}_i^0$ for each electron $i$
\begin{equation}
    \vec{m}^0_{i} = \sum_{j=1}^\nel \Gamma^{\text{e-e}}(x_{ij}, \, \vec{x}_{ij}) \odot \sigma \left( \vec{W}^\text{m} \tilde{\vec{x}}_{ij} + \vec{b}^\text{m} \right),
\end{equation}
 and compute the initial electron embeddings $\vec{h}^0_i$ as a trainable function of these messages
\begin{equation}
    \vec{h}_i^0 = \sigma \left( \vec{W}^0 \vec{m}_i^0 + \vec{b}^0 \right).
\end{equation}
The matrices $\vec{W}^\text{m}$, $\vec{W}^0$ and vectors $\vec{b}^\text{m}$, $\vec{b}^0$ are trainable parameters, $\sigma$ is an activation function which is applied elementwise and $\odot$ denotes the elementwise product.
The filter function $\Gamma^{\text{e-e}}$
\begin{align}
    \Gamma^{\text{e-e}}(x_{ij}, \, \vec{x}_{ij}) = \sigma \bigl ( \vec{W}^{\text{env}} \vec{x}_{ij} + \vec{b}\bigr) \odot \exp \bigl( -x_{ij}^2\boldsymbol{\alpha} \bigr),
\end{align}
ensures an exponential decay with a trainable vector of length-scales, $\boldsymbol{\alpha}$ and a trainable matrix $\vec{W}^{\text{env}}$. Furthermore, the input features $\tilde{\vec{x}}_{ij} = [x_{ij}, \vec{x}_{ij}, \ktwist]$ make the embedding twist dependent to allow for better transferability across twists. 

To initialize the atomic features, we first one-hot encode the nuclear charges $\vec{Z}$ into a matrix $\tilde{\vec{H}} \in \mathbb{R}^{\nnuc \times n_\text{species}}$.
We then initialize the atom embeddings $\vec{H}^0_I$ analogously to the electron embeddings, by aggregating atom-atom features for each atom $I$
\begin{equation}
\vec{H}^0_I = \sum_{J=1}^{\nnuc} \Gamma^{\text{a-a}}(x_{IJ}, \, \vec{x}_{IJ}) \odot \sigma \bigl( \vec{W}^\text{a} \tilde{\vec{H}}_J + \vec{b}^\text{a}  \bigr),
\end{equation}
using a trainable weight matrix $\vec{W}^\text{a}$ and bias vector $\vec{b}^\text{a}$.
We then incorporate electron-atom information by contracting across all electrons
\begin{align}
    \vec{H}_I^1 &= \sum_{i=1}^{\nel} \vec{e}_{iI}^0 \odot \bigl(\vec{W}^{\text{e-a}} \, \Gamma^{\text{e-a}}(x_{iI}, \, \vec{x}_{iI} ) \bigr) \\
    \vec{e}_{iI}^0 &= \sigma \bigl( \vec{h}_i^0 + \vec{H}^0_I + \vec{W}^{\text{edge}} \tilde{\vec{x}}_{iI} + \vec{b}^{\text{edge}} \bigr),
\end{align}
with $\tilde{\vec{x}}_{iI} = [x_{iI}, \vec{x}_{iI}, \ktwist]$ and trainable matrices $\vec{W}^{\text{e-a}}, \vec{W}^{\text{edge}}$ and bias $\vec{b}^{\text{edge}}$. Subsequently, the atom embeddings are updated with $L$ dense layers
\begin{align}
    \vec{H}_I^{l+1} = \sigma \bigl( \vec{W}^{l} \vec{H}_I^{l} + \vec{b}^{l} \bigr) + \vec{H}_I^{l},
\end{align}
to finally diffuse them to electron-atom embeddings $\vec{e}_{iI}$ of the form
\begin{align}
    \vec{e}_{iI} = \sigma \bigl( \vec{W}^{\text{out}_1} \vec{e}_{iI}^0 + \vec{H}_I^{L} + \vec{W}^{\text{out}_2} \vec{h}_i^0 + \vec{b}^{\text{out}}\bigr) \odot \bigl( \vec{W}^{\text{out}_3} \, \Gamma^{\text{out}}(x_{iI}, \, \vec{x}_{iI}) \bigr).
\end{align}
with trainable matrix $\vec{W}^{\text{out}_1}, \vec{W}^{\text{out}_2}, \vec{W}^{\text{out}_3}$ and trainable bias vector $\vec{b}^{\text{out}}$. For the sake of simplicity we omitted the spin dependence in this presentation of the different embedding stages.
Compared to the original Moon embedding \cite{gaoGeneralizingNeuralWave2023}, we use separate filters $\Gamma$ for the intermediate layers and the output layer, we include the twist as input feature, and omit the final aggregation step from electron-ion embeddings $\vec{e}_{iI}$ to electron embeddings $\vec{e}_i$.

\paragraph{Orbitals}
The orbital features $\tilde{\vec{c}}_{\mu J}$ are a concatenation of four different types of features. First, as proposed by Scherbela et al. \cite{scherbela2023foundation}, we rely on mean-field coefficients from a Hartree-Fock calculation. The mean-field orbitals $\phi_{\mu}$ are localized as described in \Cref{sub:orb_locaization} and expanded in periodic, atom-centered, basis functions $ b_{\eta}$
\begin{align}
    \phi_{\mu}(\vec{r}_{i}) = \sum_{I=1}^{\nnuc} \sum_{\eta=1}^\nbasis c_{I \mu , \eta}\; b_{\eta}(\vec{r}_{i} - \vec{R}_I),
\end{align}
where $\nbasis$ represents the per-atom basis set size of the Hartree-Fock calculation.
We use a periodic version of the cc-pVDZ basis set \cite{dunningGaussianBasisSets1989b} and find no strong dependence of our results on the basis set used.
Additionally, we include relative atom positions $\tilde{\vec{R}}_I$
\begin{align}
    \tilde{\vec{R}}_I = \vec{R}_I - \frac{\sum_{J=1}^{\nnuc} \vec{R}_J Z_J}{\sum_{K=1}^{\nnuc} Z_K} 
\end{align}
and analogously relative orbital positions $\tilde{\vec{R}}^{\text{orb}}_{\mu}$
\begin{align}
    \tilde{\vec{R}}^{\text{orb}}_{\mu} = \vec{R}^{\text{orb}}_{\mu} - \frac{\sum_{J=1}^{\nnuc} \vec{R}_J Z_J}{\sum_{K=1}^{\nnuc} Z_K},
\end{align}
where $\vec{R}^{\text{orb}}_{\mu}$ is the position of the localized orbital $\mu$ as outlined in \Cref{sub:orb_locaization}.
This allows the network to differentiate between different atoms and orbitals within the supercell.
As a final feature we include the twist of the system
\begin{align}
    \tilde{\vec{k}}^\text{s}_I = [\ktwist, \, \sin(\vec{R}_I \cdot \ktwist ), \cos(\vec{R}_I \cdot \ktwist )] \in \mathbb{R}^{5}.
\end{align}
The final orbital features $\tilde{\vec{c}}_{I\mu}$ are obtained as a concatenation
\begin{align}
  \tilde{\vec{c}}_{I\mu} = [\vec{c}_{I\mu}, \, \tilde{\vec{R}}_I, \, \tilde{\vec{R}}^{\text{orb}}_{\mu}, \, \tilde{\vec{k}}^\text{s}_I] \in \mathbb{R}^{\norbfeat},
\end{align}
where $\norbfeat = n_b + 11$.

\subsection{Sampling}
We use the Metropolis Hastings algorithm \cite{metropolisEquationStateCalculations1953} to draw samples $\vec{r}$ from our unnormalized density $|\Psi_\vec{\theta}|^2$.
We use Gaussian all-electron proposals $\vec{r}^\text{prop}$ of the form
\begin{equation}
    \vec{r}^\text{prop} = \vec{r} + s \vec{\delta}, \label{eq:mcmc_proposal}
\end{equation}
where $\vec{\delta}$ is drawn from a $3 \nel$-dimensional standard normal distribution. We continously adjust the stepsize $s$ to obtain a mean acceptance probability of 50\%.

When calculating properties of the hydrogen chain for different lattice constants $R$, special care must be given to the treatment of spins. 
The hydrogen chain has two phases with different arrangements of spins: In the insulating phase at large lattice constant, the ground state is antiferromagnetic, i.e. neighbouring spins prefer to be aligned antiparallel. In the metallic phase at small lattice constant, this antiferromagnetic ordering decreases and the system may even show ferromagnetic domains \cite{simonscollaborationonthemany-electronproblemGroundStatePropertiesHydrogen2020}.
Moving between these two configurations is difficult using local Monte Carlo updates as given by \Cref{eq:mcmc_proposal}, so we modify our Metropolis Hastings proposal function. In addition to moving electrons in real space, we occasionally propose moves that swap the positions of two electrons with opposite spin.
To avoid biasing our sampling towards either spin configuration, we initialize half our Monte Carlo walkers in the antiferromagnetic configuration (neighbouring electrons having opposite spin) and half our Monte Carlo walkers in a ferromagnetic configuration (all spin-up electrons in one half of the chain and all spin-down electrons in the other half).
We found that on the contrary initializing all walkers in the antiferromagnetic configuration (as might be indicated, for example, by a mean-field calculation) can cause the optimization to fall into local energy minima during wavefunction optimization. 

\subsection{Complex KFAC}
We use the Kronecker Factored Approximate Curvature (KFAC) method \cite{martens2020optimizing} to optimize the trainable parameters of our ansatz.
KFAC uses the Fisher information matrix as a metric in the space of wavefunction parameters. 
For real wavefunctions, the Fisher matrix is equivalent to the preconditioner used in the stochastic reconfiguration method~\cite{pfauInitioSolutionManyelectron2020}, but this is not the case for complex wavefunctions.
Instead, the Fubini-Study metric should be used, given by
\begin{equation}
  F_{ij} = \Re\left\{\Ev{\frac{\partial \ln \psi }{\partial \theta_i}^*\frac{\partial \ln \psi }{\partial \theta_j}}\right\}
\end{equation}
Writing the complex wavefunction in polar form, $\psi = \rho e^{i \phi}$, this becomes
\begin{equation}
  F = \Ev{
  \frac{\partial \ln \rho }{\partial \theta_i}\frac{\partial \ln \rho }{\partial \theta_j}
  + \frac{\partial \phi }{\partial \theta_i}\frac{\partial  \phi }{\partial \theta_j}
},
\end{equation}
where the first term is the Fisher information matrix and the second term is the new contribution due to the phase of the wavefunction. The second term is zero if the phase is a global constant, as is the case when the phase arises from the twist of the wavefunction only.

\subsection{Orbital localization}
\label{sub:orb_locaization}
To obtain orbital features that generalize well across system sizes, we do not use the canonical mean-field coefficients $\vec{c}$ as network inputs. Rather, we use the coefficients $\vec{c}^\text{loc}$ of maximally localized Wannier orbitals computed from $\vec{c}$.
We follow the procedure of \cite{silvestrelliMaximallyLocalizedWannier1999a} to find a unitary rotation $\vec{U}$ within the subspace spanned by the occupied orbitals.
Given a set of mean-field orbitals $\phi_{\mu}(\rvec), \mu =1,\dots,\nel$, expanded in periodic, atom-centered basis functions $b_{I\eta}(\rvec), I=1,\dots,\nnuc, \eta=1,\dots,\nbasis$, as described in \Cref{sec:architecture}, 
we compute the complex polarization matrix
\begin{equation}
    \chi_{\alpha, \nu \mu} = \int \phi_{\nu}^*(\rvec) e^{i \rvec^T \vec{G}_\alpha} \phi_{\mu}(\rvec) \dr, \quad \chi \in \mathbb{C}^{3 \times \norb \times \norb}
\end{equation}
where $\vec{G} = 2 \pi \vec{L}^{-T}$ is the matrix of reciprocal lattice vectors.
Given a unitary transformation $\vec{U} \in \mathbb{C}^{\norb \times \norb}$, the transformed polarization matrix $\hat{\vec{\chi}}$ and the corresponding localization loss $\mathcal{L}$ are given by 
\begin{align}
    &\vec{\Omega}_{\alpha \mu} = \hat{\vec{\chi}}_{\alpha, \mu \mu} = \left(\vec{U}^\dagger \chi_\alpha \vec{U}\right)_{\mu \mu} \\
    &\mathcal{L}(\vec{U}) = - || \vec{\Omega}(\vec{U}) ||^2_2,
\end{align}
where $||\cdot||_2$ denotes the L$_2$-norm.
To facilitate unconstrained optimization, we parameterize the unitary matrix $\vec{U}$ as the complex matrix exponential of a symmetrized, unconstrained complex matrix $\vec{A}$:
\begin{equation}
    \vec{U} = e^{\frac{i}{2}(\vec{A} + \vec{A}^\dagger)}.
\end{equation}
We obtain the optimal $\vec{U}^\text{loc}$, and corresponding orbital coefficients $\vec{c}^\text{loc}$ via gradient-based optimization
\begin{equation}
    \vec{U}^\text{loc} = \argmin_\vec{U} \mathcal{L}(\vec{U}), \qquad c^\text{loc}_{I\eta, \mu} = \sum_m 
 c_{I\eta, \nu} U^\text{loc}_{\nu \mu},
\end{equation}
using the Adam~\cite{kingma2017adam} optimizer.
For orthorombic supercells, the position of the Wannier center $\vec{R}^\text{orb}_{\mu}$ of the localized orbital $\mu$ can be inferred from the localized polarization matrix $\hat{\vec{\chi}}$ as 
\begin{equation}
    R^\text{orb}_{l\alpha} = - \frac{\vec{L}_{\alpha\alpha}}{2 \pi} \text{Im} \log \hat{\vec{\chi}}^\alpha_{\mu \mu}, \qquad \alpha=1\dots3,\;\;\; \mu=1\dots\norb.
\end{equation}
For other supercells we follow the generalization given in \cite{silvestrelliMaximallyLocalizedWannier1999a}.

\subsection{Observables and post-processing}
\paragraph{Twist-averaged boundary conditions}
\label{sub:tabc}
In a finite system, there are finite-size errors related to both the artificial constraint of periodicity in the supercell and the lack of correlations of longer range than the supercell.
The effects of the former on the single-particle contributions to the Hamiltonian, namely the kinetic energy, the Hartree-energy and the electron-ion interaction, can be reduced by using 
twist-averaged boundary conditions (TABC)\cite{linTwistaveragedBoundaryConditions2001,chiesaFiniteSizeErrorManyBody2006}. This means that the wavefunction obeys
\begin{equation}
    \Psi(\vec r_1, \dots, \vec r_i + \vec L_\alpha, \dots, \vec r_N)
     = e^{i \vec k \cdot \vec L_\alpha}\Psi(\vec r_1, \dots, \vec r_i, \dots, \vec r_N),
\end{equation}
where $\vec L_\alpha$ is the $\alpha$'th supercell lattice vector.
TABC are enforced by adding a position-dependent phase $e^{i \vec k_s\cdot \vec r_i}$ for each electron in the transferable atomic orbitals, as seen in \Cref{eq:orbitals_tao0} and averaging observables across a grid of twists $\vec{k}_s$ spanning first Brillouin zone.

\paragraph{Structure factor correction}
\label{sub:sfc}
In order to handle finite-size errors in the Ewald energy, we use the finite-size corrections proposed by \cite{chiesaFiniteSizeErrorManyBody2006}.
Writing the Ewald energy in terms of Fourier series, we get
\begin{equation}
  \Ev{\hat V_E} = \frac{N}{2}\left\{ v_M + \frac{1}{\Omega} \sum_{\vec G_s \neq 0} v_E(G_s)[S(\vec G_s) - 1]\right\}
  + \frac{1}{2\Omega} \sum_{\vec G_p\neq 0} v_E(G_p) \rho(\vec G_p) \rho^*(\vec G_p).
\end{equation}
Here $v_M$ is the Madelung energy, $\Omega$ is the supercell volume, $v_E(\vec k) = 4\pi/k^2$ is the Fourier transform of the Coulomb interaction, and $\vec G_s$ ($\vec G_p$) is a simulation (primitive) cell reciprocal lattice vector.
The translationally-averaged structure factor $S(\vec G_p)$ is defined by
\begin{equation}
  S(\vec G_s) = \frac{1}{N} \left [\ev{\hat \rho(\vec G_s) \hat \rho^*(\vec G_s)} - \ev{\hat \rho(\vec G_s)}\ev{\hat \rho^*(\vec G_s)}\right],
\end{equation}
where $\hat \rho(\vec G_s) = \sum_i \exp\left(-i \vec G_s \cdot \vec r_i\right)$ is the Fourier representation of the operator for the electron density.
The structure factor converges fairly rapidly with supercell size, so we can assume that $S_\Omega(\vec k) \approx S_\infty(\vec k)$.
In this limit, the largest contribution to the error is the omission of the $G_s = 0$ term in the first sum.
In cubic systems,  we have $S(\vec k) \propto \eta k^2 + O(k^4)$, with odd terms missing due to inversion symmetry, and the $\vec k \to 0$ limit of $S(\vec k) v_E(k)$ is well defined.
As such, to a first approximation, the Ewald finite-size error is given by
\begin{equation}
  \Delta V_E \approx \frac{N}{2\Omega} \lim_{k\to 0} v_E(k)S(\vec k) = \frac{4\pi N}{2\Omega} \lim_{k\to 0} \frac{S(\vec k)}{k^2}.
\end{equation}
Sampling $S(\vec G_s)$ at supercell reciprocal lattice vectors $\vec G_s$, we approximate the limit $k\to 0$ by fitting the function
\begin{equation}
  S(k) \approx f(k) = 1 - e^{-a_0 k^2 - a_1 k^4},
\end{equation}
with $a_0$ and $a_1$ greater than zero.
The form of the fit ensures that $S(k)$ has the correct $k^2$ behavior at small $k$ and that  $\lim_{k \to \infty} S(\vec k) = 1$. The finite-size correction $\Delta V_E \approx 4\pi N a_0 / 2 \Omega$.

\paragraph{Zero-point vibrational energy for graphene}
\label{sub:zpve}
To estimate the zero-point vibrational energy (ZPVE) contribution for graphene, we obtained the phonon density of states $D(\omega)$ calculated within DFT at the PBE \cite{perdewGeneralizedGradientApproximation1996a} level from \cite{dieryNatureLocalizedPhonon2018}.
The ZPVE energy per primitive cell, $E_\text{ZPVE}$, is then given as 
\begin{equation}
E_\text{ZPVE} = \frac{3 N^\text{prim}_\text{atoms}}{\int D(\omega) \text{d}\omega}\int D(\omega) \frac{1}{2}\hbar \omega \text{d}\omega,
\end{equation}
where $N^\text{prim}_\text{atoms}=2$ is the number of atoms per primitive unit cell of graphene.
This yields a ZPVE of 12.8~mHa / primitive cell.

%
%
\section{Code availability}
All code is available on our github repository \url{https://github.com/mdsunivie/deeperwin}.
\section{Acknowledgements}
We acknowledge helpful discussions with Ji Chen, Xiang Li, Tony Lou, and Gunnar Arctaedius.
We gratefully acknowledge financial support from the following grants: Austrian Science Fund FWF Project I 3403 (P.G.), WWTF-ICT19-041 (L.G.) and Aker Scholarship (H.S.).
The computational results have been partially achieved using the Vienna Scientific Cluster and Leonardo (Project L-AUT 005).
HS gratefully acknowledges the Gauss Centre for Supercomputing e.V. (www.gauss-centre.eu) for providing computing time through the John von Neumann Institute for Computing (NIC) on the GCS Supercomputer JUWELS at Jülich Supercomputing Centre (JSC);
the HPC RIVR consortium and EuroHPC JU for resources on the Vega high performance computing system at IZUM, the Institute of Information Science in Maribor;
and the UK Engineering and Physical Sciences Research Council for resources on the Baskerville Tier 2 HPC service.
Baskerville was funded by the EPSRC and UKRI through the World Class Labs scheme (EP/T022221/1) and the Digital Research Infrastructure programme (EP/W032244/1) and is operated by Advanced Research Computing at the University of Birmingham.
The funders had no role in study design, data collection and analysis, decision to publish or preparation of the manuscript.
\section{Author Contributions Statement}
HS proposed the idea of shared twist averaging; MS and LG proposed concrete architecture and approach.
LG, MS and HS jointly worked on implementation: LG worked on the periodic Hamiltonian and contributed to the model. MS implemented/extended the model, orbital localization and contributed to the mean-field orbitals. HS extended KFAC to complex wavefunctions, implemented the evaluation of the mean-field orbitals and contributed to the periodic Hamiltonian.
MS designed and ran experiments on the H chains. LG designed and ran experiments for LiH and graphene with support from HS and MS.
LG, MS, HS and WMCF jointly wrote the paper with supervision and funding from PG and WMCF.

\section{Competing Interests Statement}
The authors declare no competing interests.

\newpage
\printbibliography

\newpage
\setcounter{equation}{0}
\setcounter{figure}{0}
\setcounter{table}{0}
\setcounter{page}{1}
\setcounter{section}{-1}\stepcounter{section}
\renewcommand{\theequation}{S\arabic{equation}}
\renewcommand{\thefigure}{S\arabic{figure}}
\renewcommand{\thesection}{S\arabic{section}}

\begin{center}
\textbf{\Large Supplementary Information for\\Transferable Neural Wavefunctions for Solids}
\end{center}
\section{Scaling of compute cost with system size}
\label{sec:scaling_si}
\begin{figure}[!htpb]
    \centering
    \includegraphics[width=0.5\textwidth]{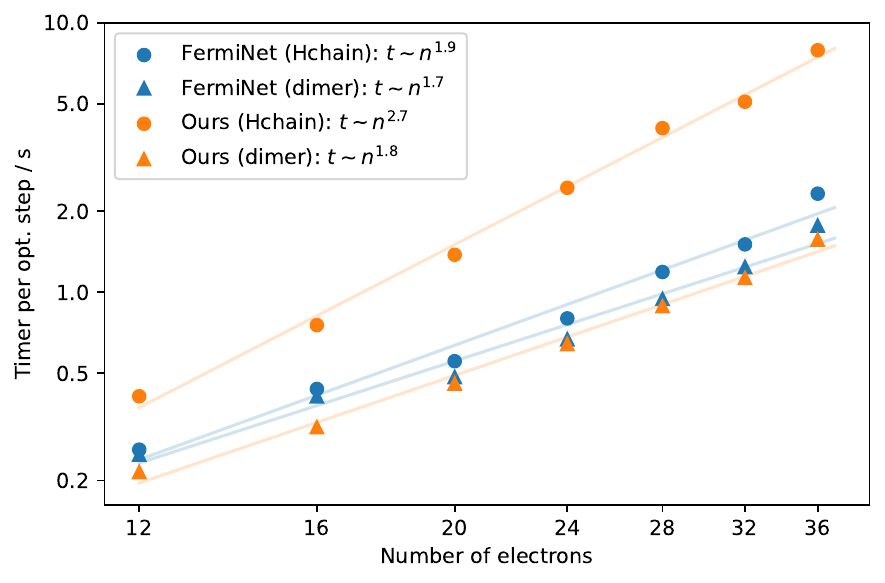}
    \caption{\textbf{Scaling of computational cost}: Markers correspond to measured timings, lines correspond to least-square fits of power-laws, with the exponents denoted in the legend.}
    \label{fig:scaling}
\end{figure}

Computing the orbital matrix $\Phi_{ik}$ in our ansatz for each electron $i$ and orbital $k$ requires a sum over all nuclei $J$. Since the number of orbitals and electrons are equal to $\nel$ and the number of nuclei $\nnuc$ is in a worst-case also equal to $\nel$, materializing this matrix has a worst-case scaling of $\mathcal{O}(\nel^3)$.
This is in contrast to other approaches such as FermiNet, where this matrix is not given as a sum over nuclei and thus only scales as $\mathcal{O}(\nel^2)$.
Because scaling in the limit of $\nel \to \infty$ is in either case dominated by the evaluation of the determinant, which scales as $\mathcal{O}(\nel^3)$, this does not impact the overall scaling of the method, but can lead to different empirical scaling.
\Cref{fig:scaling} depicts median run-time per optimization step for FermiNet (our implementation) and our approach.
We compare timings on chains of Hydrogen atoms of increasing length and dimers of increasing nuclear charge. The former depicts the worst case for our method, the latter is close to the best case.
All timings are obtained on 2 A100-GPUs using a batch-size of 512 and 8 determinants.

\section{Limited expressiveness of electron-wise embeddings}
\label{sec:limited_expressiveness}
Prior work has relied on embeddings $\vec{h}_i$ for each electron $i$ to construct correlated orbitals. 
We show that requiring the following three reasonable symmetries already overly constrains which embeddings $\vec{h}_i$ (and consequently orbitals $\Phi$) can be represented:

Invariance wrt. to continuous translation of \emph{all} particles:
\begin{equation}
\label{eq:invariance_cont_trans}
h(\rvec_1 + \vec{\delta}, \dots\, \rvec_\nel + \vec{\delta}, \Rvec_1 + \vec{\delta},\dots, \Rvec_\nnuc+ \vec{\delta}) = h(\rvec_1,\dots,\rvec_\nel, \Rvec_1,\dots, \Rvec_\nnuc).
\end{equation}

Invariance wrt. to permutation of nuclei of same charge Z:
\begin{equation}
\label{eq:invariance_nuclei_permutation}
h(\rvec_1,\dots,\rvec_\nel, \Rvec_1,\dots,\Rvec_I,\dots\Rvec_J,\dots \Rvec_\nnuc) = h(\rvec_1,\dots,\rvec_\nel, \Rvec_1,\dots,\Rvec_J,\dots\Rvec_I,\dots \Rvec_\nnuc).
\end{equation}

Invariance wrt. to translation of any particle by a supercell lattice vector $\latsc$:
\begin{equation}
\label{eq:invariance_supercell_translation}
h(\rvec_1 + \latsc,\dots,\rvec_\nel, \Rvec_1,\dots \Rvec_\nnuc) = h(\rvec_1,\dots,\rvec_\nel, \Rvec_1, \Rvec_\nnuc).
\end{equation}

To demonstrate the problem consider a simplified 1D example with a single electron and a supercell consisting of $N$ primitive cells, with lattice consant $a$, each containing a single nucleus.
The coordinates $R_J$ of all nuclei in the supercell are thus given by
\begin{equation}
    R_J = Ja.
\end{equation}

An embedding satisfying the invariances \cref{eq:invariance_cont_trans}, \cref{eq:invariance_nuclei_permutation}, \cref{eq:invariance_supercell_translation} is given by any permutation invariant function $h$
\begin{align}
    h &= h\left((\omega(r-R_1), \dots, \omega(r-R_\nnuc)\right) \label{eq:embedding_1d}\\
     &= h(\omega(r-a), \omega(r-2a), \dots, \omega(r-Na)),
\end{align}
where $\omega$ computes input features that are periodic in the supercell
\begin{equation}
    \omega(x) = \omega(x + L_\text{sc}) = \omega(x + Na).
\end{equation}
Here using distances $r-R$ automatically enforces \cref{eq:invariance_cont_trans} and using periodic versions $\omega(r-R)$ of these distances automatically enforces \cref{eq:invariance_supercell_translation}.

For this system, any embedding following the structure in \cref{eq:embedding_1d} is necessarily not only invariant under translations of electrons by a \emph{supercell} lattice vector, but also invariant under translation of electrons by a \emph{primitive} lattice vector.
\begin{align}
h^\text{shifted}_\text{prim} &= h(r+a, R_1, \dots, R_\nnuc) \\
&=h\left((\omega(r+a-a), \omega(r+a - 2a),\dots, \omega(r+a-Na)\right)\\
&\stackrel{\text{\ref{eq:invariance_nuclei_permutation}}}{=}h\left((\omega(r-a), \omega(r - 2a),\dots, \omega(r)\right)\\
&\stackrel{\text{\ref{eq:invariance_supercell_translation}}}{=}h\left((\omega(r-a), \omega(r - 2a),\dots, \omega(r-Na)\right)\\
&=h(r, R_1, \dots, R_\nnuc) = h^\text{orig}.
\end{align}
Therefore using electron embeddings with these symmetries allows only representation of orbitals that are periodic on the primitive lattice.
This excludes many relevant functions such as localized orbitals and prevents the network from representing long-range correlations.
To break this unwanted symmetry there are at least three options: 
\begin{itemize}
    \item Break invariance with respect to permutation of nuclei. Non-transferable ansätze such as FermiNet \cite{pfauInitioSolutionManyelectron2020}, PsiFormer \cite{vonglehnSelfAttentionAnsatzAbinitio2022} or DeepSolid \cite{li_ab_solids_2022} all break this permutation invariance. Since these approaches are only ever trained on a single system (and thus a single permutation of nuclei) this poses no issue there, but prevents efficient generalization to permuted, but physically identical systems.
    \item Break supercell lattice translational symmetry. For gas-phase calculations there is no periodicity and thus existing transferable approaches \cite{gaoGeneralizingNeuralWave2023, scherbela2023foundation, scherbela2023variationalOnABudget} do not face this issue. For periodic systems however periodicity is required to be able to enforce boundary conditions.
    \item Use permutation \emph{equivariant} electron-ion embeddings instead of permutation \emph{invariant} electron embeddings, as done in this work.
\end{itemize}

\section{Hyperparameters}

A detailed description of the hyperparameter used in this work can be found below (cf. \cref{table:hyperparams}). For optimization we rely on the second-order method KFAC \cite{martens2020optimizing} and use their Python implementation \cite{kfac-jax2022github_b}.  
\begin{table*}[!htb]
\caption{Hyperparameter settings used in this work \label{table:hyperparams}}
\begin{tabularx}{\textwidth}{lXc}
\toprule
\multirow{2}{*}{
\textbf{HF-pre-training}
}
		 &Pre-training basis set & cc-pVDZ\\
		 &Pre-training steps per geometry & 100-500\\
\midrule
\multirow{7}{*}{ \textbf{Embedding}}
    &Envelope power $\gamma$ & 2\\
    &Uniform initialization of envelope scaling $\alpha$ & 8-10\\
    &El-el hidden dimension    & 32\\ 
    &El-Ion hidden dimension    & 128\\
    &Ion hidden dimension & 128\\
    &\textnumero\, hidden layers of ion embedding & 3\\
    &Final el-ion embedding dimension $e_{iI}$ & 64\\
\midrule
\multirow{9}{*}{\shortstack[l]{
\textbf{Transferable}\\
\textbf{atomic orbitals}
}}
	& \textnumero\, determinants $\ndet$   & 8\\ 
    & Basis set for orbital descriptor & cc-pVDZ \\
    & \textnumero\, hidden layers of $f^W$ & 2\\
    & Hidden dimension of $f^W$      & 128\\     
    & \textnumero\, hidden layers of $f^a$ & 2\\
    & Hidden dimension of $f^a$      & 32\\   
	& Activation function & ReLU\\
    & Residual connection & True \\
    & Layer Norm & True\\
\midrule
\multirow{2}{*}{ \textbf{Jastrow factor}}
    & \textnumero\, hidden layers of MLP  & 2\\
    & Hidden dimension of MLP      & 40\\   
\midrule
\multirow{3}{*}{\shortstack[l]{
\textbf{Markov Chain}\\
\textbf{Monte Carlo}
}}
		 &\textnumero\, walkers                        & $2048$	\\
		 &\textnumero\, decorrelation steps            & 20	\\
		 &Target acceptance prob.         & 50\%	\\
\midrule
\multirow{7}{*}{\shortstack[l]{
\textbf{Variational}\\
\textbf{optimization}
}}
		 &Optimizer		                        & KFAC\\
		 &Damping 	                            & $1-3 \times 10^{-3}$ 	\\
		 &Norm constraint	                    & $1 \times 10^{-3}$ 	\\
		 &Batch size 	                        & $2048$	\\
		 &Initial learning rate $\text{lr}_0$	& $0.1-0.3$	\\
		 &Learning rate decay   	            & $\text{lr}(t) = \text{lr}_0 (1+t/6000)^{-1}$ 	\\
		 &Optimization steps                    &100,000 - 200,000\\
\midrule
\multirow{2}{*}{\shortstack[l]{
\textbf{Changes for}\\
\textbf{Reuse}
}}
    &Initial learning rate $\text{lr}_0$	& $0.05$	\\
	&Learning rate decay   	            & $\text{lr}(t) = \text{lr}_0 (1+t/6000)^{-1}$ 	\\
	&Optimization steps                    &0 - 10,000\\
\bottomrule
\end{tabularx}
\end{table*}

\newpage

\section{Total energies}
\label{sec:si_total_energies}
For better comparison we add the total energies of the results in \Cref{tab:graphene_tabc_results} and \Cref{fig:lih_twist_avg_pes}. For Graphene (see \Cref{sec:graphene}) we state the total energies per primitive cell for each twist plus the combination of structure-factor-based finite size corrections and ZPVE in \Cref{tab:graphene_k_resolved_results}. For LiH (see \Cref{sec:lih}) we state the twist averaged energies per primitive cell for each lattice constant, separately depicting structure-factor-based finite size correction and ZPVE in \Cref{tab:lih_tabc_results}

\begin{table}[htb]
    \centering
    \caption{
    \textbf{Total energies of graphene} in Hartrees. 
    The energies depict the total energies per primitive cell by sequentially adding structure-factor-based finite-size correction (SFC) and ZPVE. The systems represent the symmetry-inequivalent twists for the $12 \times 12$ Monkhorst-Pack grid.}
    \begin{tabular}{lllll}
    \textbf{Twists} &\textbf{Total Energy} & \makecell[l]{\textbf{Total energy} \\ \textbf{+ SFC}}  & \makecell[l]{\textbf{Total energy} \\ \textbf{+ SFC} \\ \textbf{+ ZPVE}}  \\ 
    \toprule
    $\vec{k}=(0.33, 0.33)$ & $-76.2572$ & $-76.2543$ & $-76.2415$\\ 
    $\vec{k}=(0, -0.17)$ & $-76.2002$ & $-76.1972$ & $-76.1844$\\ 
    $\vec{k}=(-0.08, -0.42)$ & $-76.2679$ & $-76.2650$ & $-76.2522$\\ 
    $\vec{k}=(-0.08, -0.33)$ & $-76.2470$ & $-76.2441$ & $-76.2313$\\ 
    $\vec{k}=(-0.08, -0.25)$ & $-76.2190$ & $-76.2160$ & $-76.2032$\\ 
    $\vec{k}=(-0.17, -0.42)$ & $-76.2615$ & $-76.2586$ & $-76.2458$\\ 
    $\vec{k}=(0.67, 0.33)$ & $-76.2590$ & $-76.2560$ & $-76.2432$\\ 
    $\vec{k}=(-0.17, -0.58)$ & $-76.2775$ & $-76.2746$ & $-76.2618$\\ 
    $\vec{k}=(-0.08, -0.17)$ & $-76.1914$ & $-76.1884$ & $-76.1756$\\ 
    $\vec{k}=(-0.08, -0.50)$ & $-76.2786$ & $-76.2757$ & $-76.2629$\\ 
    $\vec{k}=(0, 0)$ & $-76.1572$ & $-76.1542$ & $-76.1414$\\ 
    $\vec{k}=(0, -0.25)$ & $-76.2308$ & $-76.2278$ & $-76.2150$\\ 
    $\vec{k}=(-0.17, -0.33)$ & $-76.2407$ & $-76.2377$ & $-76.2249$\\ 
    $\vec{k}=(-0.25, -0.50)$ & $-76.2688$ & $-76.2658$ & $-76.2530$\\ 
    $\vec{k}=(0, -0.08)$ & $-76.1723$ & $-76.1693$ & $-76.1565$\\ 
    $\vec{k}=(0, -0.50)$ & $-76.2787$ & $-76.2758$ & $-76.2630$\\ 
    $\vec{k}=(0, -0.42)$ & $-76.2737$ & $-76.2708$ & $-76.2580$\\ 
    $\vec{k}=(-0.25, -0.58)$ & $-76.2728$ & $-76.2699$ & $-76.2571$\\ 
    $\vec{k}=(-0.17, -0.50)$ & $-76.2740$ & $-76.2710$ & $-76.2582$\\ 
    \midrule
     Averaged $3 \times 3$& $-76.2465$ & $-76.2435$ & $-76.2307$ \\
     Averaged $12 \times 12$& $-76.2503$ & $-76.2473$ & $-76.2345$ \\
    \midrule
    \bottomrule
    \label{tab:graphene_k_resolved_results}
    \end{tabular}
\end{table}

\begin{table}[htb]
    \centering
    \caption{
    \textbf{Total energies of LiH} in Hartrees. 
    The energies depict the twist averaged total energies per primitive cell by sequentially adding structure-factor-based finite-size correction (SFC) and ZPVE. For the twist averaging we use a $5 \times 5 \times 5$ Monkhorst-Pack grid per lattice constant. The energies accompanies the \Cref{fig:lih_twist_avg_pes}.}
    \begin{tabular}{llllll}
    \makecell[l]{\textbf{Supercell} \\ \textbf{size}} & \makecell[l]{\textbf{Lattice} \\ \textbf{constant / $a_0$}} &\textbf{Total Energy} & \makecell[l]{\textbf{Total energy} \\ \textbf{+ SFC}}  & \makecell[l]{\textbf{Total energy} \\ \textbf{+ SFC} \\ \textbf{+ ZPVE}}  \\ 
    \toprule
    \multirow{8}{*}{$2 \times 2 \times 2$}  & $6.4$ & $-8.1654$ & $-8.1462$ & $-8.1338$ \\ 
                                            & $6.8$ & $-8.1752$ & $-8.1574$ & $-8.1464$ \\ 
                                            & $7.2$ & $-8.1792$ & $-8.1626$ & $-8.1529$ \\ 
                                            & $7.6$ & $-8.1789$ & $-8.1636$ & $-8.1554$ \\ 
                                            & $7.9$ & $-8.1759$ & $-8.1618$ & $-8.1545$ \\ 
                                            & $8.3$ & $-8.1709$ & $-8.1580$ & $-8.1522$ \\ 
                                            & $8.7$ & $-8.1646$ & $-8.1527$ & $-8.1482$ \\ 
                                            & $9.1$ & $-8.1574$ & $-8.1467$ & $-8.1436$ \\ 
    \midrule
    $3 \times 3 \times 3$ & $7.674$ & $-8.1644$ & $-8.1604$ & $-8.1524$ \\
    \midrule
    \bottomrule
    \label{tab:lih_tabc_results}
    \end{tabular}
\end{table}

%
%
\end{document}